\documentstyle[12pt]{article}
\begin{document}

\large
\begin{center}

{\bf Deep Inelastic Scattering from
     Polarized Deuterons$^{\dagger}$}

\vspace*{1cm}

\normalsize

W.Melnitchouk

{\em Institute for Theoretical Physics,
University of Regensburg,
D-93040 Regensburg, Germany}

\vspace*{0.5cm}

G.Piller and A.W.Thomas

{\em Department of Physics and Mathematical Physics,
     University of Adelaide,
     South Australia, 5005,
     Australia}
\end{center}

\begin{abstract}
The spin-dependent structure function of the deuteron, $g_{1D}$,
is calculated within a covariant framework.
The off-shell structure of the bound nucleon gives corrections
to the convolution model at a level of half a percent for $x$
below 0.7, increasing to more than five percent at larger $x$.
Overall, the dominant source of error comes from the lack of
knowledge associated with the deuteron $D$-state, which may
introduce an uncertainty in the neutron spin structure function,
$g_{1n}$, extracted from deuterium data of up to ten percent
for $x$ around 0.2.
\end{abstract}

\normalsize

\vspace*{1.5cm}

ADP-94-16/T157

TPR-94-22

To appear in Physics Letters B.

\vspace*{2.5cm}

$^{\dagger}$ Work supported in part by BMFT grant 06 OR 735.

\newpage

There is currently much discussion about the interpretation
of the results from the Spin Muon Collaboration \cite{SMCD} and
SLAC E142 \cite{E142} experiments on the
polarized deuteron and helium structure functions.
Of particular interest is the neutron structure function,
$g_{1n}$.
A precise knowledge of this is necessary to verify
the Bjorken sum rule --- when combined with the previously
measured proton structure function $g_{1p}$
\cite{SLAC,EMC,SMCP,E143P}.
In addition, an independent measurement of $g_{1n}$ is important
for determining the flavor singlet combination of polarized quark
distributions --- the first moment of which, in the naive parton model,
is just the fraction of the spin of the nucleon carried by quarks.

Before the nuclear data can be used for these purposes,
it is essential to account for any nuclear effects
that may arise when extracting $g_{1N}$ from $g_{1A}$.
Nuclear corrections to polarized deuteron and helium structure
functions, such as those due to Fermi motion, shadowing or
meson-exchange currents, have been considered by several authors
\cite{KH,FS,WOL,CIOFI,TOK,KAP}.
Early attempts \cite{FS} to describe nuclear effects in the
deuteron, based on time-ordered perturbation theory
in the infinite momentum frame, suffered from
the problem that the deuteron wavefunctions
are not known in this frame.
Subsequent analyses \cite{WOL,CIOFI}
all adopted the so-called convolution approach, in which
the nuclear structure function is a one-dimensional
convolution of the structure function of a free nucleon
with the nucleon momentum distribution in the nucleus.
Efforts to incorporate relativistic effects in the deuteron
were made in Refs.\cite{TOK,KAP}, however also within the
confines of the convolution model.

In contrast with the earlier work,
in this letter we demonstrate that in a covariant
treatment, inclusion of the full off-mass-shell structure
of bound nucleons leads to a breakdown of convolution for
spin-dependent nuclear structure functions.
A convolution component can, however, be extracted from the
full result by selectively taking on-shell limits for the
bound nucleon structure function, and neglecting relativistic
components of the nuclear wavefunction.
Off-shell corrections to the spin-averaged structure
function of the deuteron, $F_{2D}$, were examined in Ref.\cite{MSTD},
and found to be about 1-2\% for $x < 0.9$
(c.f. heavy nuclei or nuclear matter, where the off-shell effects
can be considerably larger \cite{MST,KPW}).
Since the absolute values of $g_{1D}$ and $g_{1n}$
are considerably smaller than $F_{2D}$ or $F_{2n}$,
we may expect nuclear corrections to be
of greater relative importance for the spin-dependent
structure functions.
It is of some importance therefore that the issue of
off-shell effects in $g_{1D}$ be seriously addressed.

In the present analysis we restrict ourselves to
the valence component of polarized structure functions,
thereby avoiding confronting the issue of the axial anomaly \cite{U1},
which could divert attention from our main interest.
Within the impulse approximation,
deep inelastic
scattering from a polarized deuteron
is described as a two-step process,
in terms of the virtual photon--nucleon interaction,
parametrized by the truncated antisymmetric nucleon tensor
$\widehat G_{\mu\nu}(p,q)$,
and the polarized nucleon--deuteron scattering amplitude,
$\widehat A(P,p,S)$. The antisymmetric part of the deuteron
hadronic tensor can then be written as:
\begin{eqnarray}
M_D \,W_{\mu\nu}^{D}(P,S,q)
&\equiv& i { M_D \over P\cdot q }
\epsilon_{\mu \nu \alpha \beta}\ q^{\alpha}
\left( S^{\beta}\ g_{1D}(P,q)\
    +\ \left( S^{\beta} - { S\cdot q \over P\cdot q } P^{\beta}
       \right) g_{2D}(P,q)
\right)                             \nonumber\\
&=& \int \frac{ d^4 p }{ (2\pi)^4 }
    {\rm Tr} \left[ \widehat{A}(P,S,p)\ \
                    \widehat{G}_{\mu\nu}(p,q)
             \right]\
    2\pi \delta\left( (P-p)^2 - M^2 \right),    \label{Wmunu}
\end{eqnarray}
where $P$, $p$ and $q$ are the deuteron, off-shell nucleon
and photon momenta, respectively, and $M_D$ is the
deuteron mass\footnote{
For a discussion of some problems associated with using the
impulse approximation for $g_{2A}$ see Ref.\cite{MS}.}.
The vector $S^{\alpha}$ ($S^2=-1, P\cdot S=0$) is defined
in terms of deuteron polarization vectors $\varepsilon_{\alpha}^m$
such that \cite{JHM}
$S^{\alpha}(m)
\equiv  -i\ \epsilon^{\alpha\beta\lambda\rho}\
            \varepsilon^{m *}_{\beta}\ \varepsilon^{m}_{\lambda}\
            P_{\rho} / M_D,
$
where $m=0,\pm1$ is the spin projection.

In analyzing nucleon off-shell effects, it will be convenient
to expand the truncated nucleon tensor
$\widehat G_{\mu\nu}$ in terms of independent
basis tensors:
\begin{eqnarray}                                \label{Whatex}
\widehat{G}_{\mu\nu}(p,q)
&=& i \epsilon_{\mu \nu \alpha \beta}\ q^{\alpha}
\left( p^{\beta} \left( \not\!p \gamma_5\, G_{(p)}\
                     +\ \not\!q \gamma_5\, G_{(q)}
                 \right)
    +\ \gamma^{\beta} \gamma_5\, G_{(\gamma)}\
    +\ \cdots
\right),
\end{eqnarray}
where the coefficients $G_{(i)}$ are scalar functions
of $p$ and $q$,
and the dots ($\dots$) represent
terms that do not contribute to the $g_1$
structure function in the Bjorken limit,
as well as those that vanish in the massless quark limit
(which we use throughout).

The polarized structure function, $g_{1D}$, can be
extracted from $W_{\mu\nu}^D$ by considering the
polarization combination ($m$=+1) -- ($m$=--1).
Using the fact that
$ \left( \varepsilon_{\alpha}^{+}\ \varepsilon_{\beta}^{+*}\
      -\ \varepsilon_{\alpha}^{-}\ \varepsilon_{\beta}^{-*}\
  \right)
= -i\ \epsilon_{\lambda\rho\alpha\beta}\ P^{\lambda} S^{\rho} / M_D $\
(since $\varepsilon_{\alpha}^{+*} = - \varepsilon_{\alpha}^{-}$),
the deuteron--nucleon amplitude $\widehat{A}$ can be written:
\begin{eqnarray}
\widehat{A}
&=& - {i \over 2 M_D} \epsilon_{\lambda\rho\alpha\beta}\
                      P^{\lambda} S^{\rho}\
    (\not\!p - M)^{-1}\
    \Gamma^{\alpha}(p)\
    (\not\!P - \not\!p - M)\
    \overline{\Gamma}^{\beta}(p)\
    (\not\!p - M)^{-1},
\end{eqnarray}
where $\Gamma^{\alpha}(p)$ is the relativistic deuteron--nucleon
vertex function \cite{BG}.
In the massless quark limit only the pseudovector
components of $\widehat A$ are relevant:
$\widehat A \equiv \gamma_5 \gamma_{\lambda} {\cal A}^{\lambda}$.
In terms of ${\cal A}^{\lambda}$ and $G_{(i)}$,
the $g_{1D}$ structure function (per nucleon) is:
\begin{eqnarray}
\hspace*{-0.2cm}
g_{1D}(x)
&=& {P \cdot q \over 4 \pi^2\ M_D\ S \cdot q}
    \int dy\ dp^2
    \left( {\cal A} \cdot q\
           \left( p \cdot q\ G_{(q)}\ +\ G_{(\gamma)}
           \right)\
    \right.                             \nonumber\\
& & \hspace*{4cm}
    \left.  +\ p \cdot q\ {\cal A} \cdot p\ G_{(p)}
    \right),                            \label{g1Dfull}
\end{eqnarray}
where $x = Q^2 / 2 P\cdot q$ is the Bjorken scaling variable
and $y = p \cdot q / P \cdot q$ is the light-cone
fraction of the deuteron momentum carried by the struck
nucleon.

The analogous hadronic tensor for an on-shell nucleon is
obtained by tracing $\widehat G_{\mu\nu}$ with the
nucleon spin-energy projector:
\begin{eqnarray}
2 M\ W_{\mu\nu}^{N}(p,q)
&=& {\rm Tr}
    \left[ \left( \not\!p + M \right)\
           \left( 1 + \gamma_5~{\not\!s} \right)\
           \widehat G_{\mu\nu}(p,q)
    \right],
\end{eqnarray}
where $s$ is the nucleon polarization vector,
and by setting $p^2=M^2$ and $y=1$ in the
coefficients $G_{(i)}$.
The structure function $g_{1N}$
can then be identified as\
$g_{1N}(x) = \widetilde g_{1N}(x,p^2=M^2)$,
where
\begin{eqnarray}
\widetilde g_{1N} \left( {x \over y}, p^2 \right)
&=& 2 p \cdot q\ \left( p \cdot q\ G_{(q)} + G_{(\gamma)} \right).
                                                        \label{g1N}
\end{eqnarray}
The presence of the ${G}_{(p)}$ term in Eq.(\ref{g1Dfull})
(which does not appear in $g_{1N}$) means that simple factorization
of the fully relativistic $g_{1D}$ into nuclear
(${\cal A}^{\lambda}$) and nucleon ($G_{(i)}$) parts
is not strictly possible.
This however would be required for convolution, as assumed in
Refs.\cite{WOL,CIOFI,TOK}.
Nonetheless, by writing ${\cal A}^{\lambda}$ in terms
of relativistic deuteron wavefunctions, as calculated for example
in \cite{BG} (see also \cite{RELDWFN}), we can show that
the $G_{(p)}$ term is of order $(v/c)^3$
compared with the first term in the integrand in Eq.(\ref{g1Dfull}).
Indeed, all of the non-factorizable corrections to
convolution can be shown to be of higher orders in $v/c$.
This is easily seen by separating the $``+"$ component of
${\cal A}^{\lambda}$ into an ``on-shell" part,
which is proportional to the non-relativistic
deuteron wavefunctions (see Eq.(\ref{PsiNR}) below),
and an off-shell component,
${\cal A}^+ \equiv {\cal A}^+_{ON} + {\cal A}^+_{OFF}$,
where (in the deuteron rest frame):
\begin{eqnarray}
{\cal A}^+_{ON}
&=& 2 \pi^2 M_D M
    \left[ u^2 + \left( 1 - 3 \cos^2\theta \right) {u w\over\sqrt{2}}
               - \left( 1 - {3\over 2} \cos^2\theta \right) w^2
    \right],                                    \label{A+con}
\end{eqnarray}
and
\begin{eqnarray}
{\cal A}^+_{OFF}
&=& 2 \pi^2 M_D M
    \left[
    \left( {p_z \over M}
            - \left( 1 - { E_p \over M } \right) \cos^2\theta
    \right)
    \left( u - { w \over \sqrt{2} } \right)^2
    \right.                                                     \nonumber\\
& & -\ {3 \over 2}
       \left( {p_z \over M}
            - { E_p \over M } \cos^2\theta
       \right) v_t^2\
 +\ {3 \over \sqrt{2}} \left( 1 - \cos^2\theta \right) v_s v_t  \nonumber\\
& & +\ \sqrt{3} \left( \cos\theta
                     - {|{\bf p}| \over 2 M}
                       \left( 1 - \cos^2\theta \right)
                \right) u v_t\                                  \nonumber\\
& & -\ \sqrt{3} \left( \cos\theta
                     - {|{\bf p}| \over M}
                       \left( 1 - \cos^2\theta \right)
                \right) w v_t\                                 \nonumber\\
& & \left.
    +\ {\sqrt{3} |{\bf p}| \over M} \left( 1 - \cos^2\theta \right)
       \left( u - {w\over\sqrt{2}} \right) v_s
    \right],                                            \label{A+off}
\end{eqnarray}
with
$p_z = |{\bf p}| \cos\theta$,
$E_p = \sqrt{ M^2 + {\bf p}^2 }$ and
$\cos\theta = (y M_D - p_0) / |{\bf p}|$.
In Eqs.(\ref{A+con}) and (\ref{A+off}) $u$ and $w$ correspond to
the $S$- and $D$-state deuteron wavefunctions,
while $v_t$ and $v_s$ are the relativistic
triplet and singlet $P$-state contributions.
(In our numerical calculation we use the model of
Ref.\cite{BG} with wavefunctions
corresponding to the pseudovector $\pi NN$ interaction.)

Using Eqs.(\ref{g1N})--(\ref{A+off}) we can then decompose
$g_{1D}$ into a convolution component plus off-shell corrections:
\begin{eqnarray}
g_{1D}(x)
= \int_x^1\ {dy \over y } \Delta f(y)\ g_{1N}\left({x\over y}\right)\
 +\ \delta^{(N)} g_{1D}(x)\
 +\ \delta^{({\cal A})} g_{1D}(x)\
 +\ \delta^{(G)} g_{1D}(x),             \label{comps}
\end{eqnarray}
where \cite{WOL}
\begin{eqnarray}
\Delta f(y)
= \int d^4p\
  \left( 1 + { p_z \over M } \right)\
  \Delta S(p)\
  \delta \left( y - \frac{p^+}{M_D} \right),            \label{Dfy}
\end{eqnarray}
can be identified with the difference of probabilities
to find a nucleon in the deuteron with momentum fraction
$y$ and spin parallel and antiparallel to that of the deuteron.
For a deuteron with polarization $m=+1$,
$\Delta S(p)$ corresponds to the spectral function:
\begin{eqnarray}
\Delta S(p)
= \Psi_{+1}^{\dagger}(p)\
  \widehat S_z\
  \Psi_{+1}(p)\ \
  \delta\left( p_0 - M_D + E_p \right),         \label{DSp}
\end{eqnarray}
where $\Psi_m(p)$ is the usual (normalized) non-relativistic deuteron
wavefunction, and $\widehat S_z$ is the $z$ component
of the nucleon spin operator.
In terms of $\Psi_m(p)$, ${\cal A}^+_{ON}$ can be written:
\begin{eqnarray}
{\cal A}^+_{ON}
= 8 \pi^3 M_D M \ {\cal N}\ \
  \Psi_{+1}^{\dagger}(p)\
  \widehat S_z\
  \Psi_{+1}(p),                         \label{PsiNR}
\end{eqnarray}
where
${\cal N} = \int d|{\bf p}|\ {\bf p}^2\ (u^2 + w^2)$
ensures that the normalization agrees with that of the relativistic
calculation.
The function $\Delta f(y)$ then satisfies
$ \int_0^1 dy\ \Delta f(y)
= 1 - 3/2\ \omega_D $,
where
$\omega_D = \int d|{\bf p}|\ {\bf p}^2\ w^2\ /\ {\cal N}$
is the non-relativistic $D$-state probability.

We should point out that the definition of the convolution
component in Eq.(\ref{comps}) is not unique.
For example, some authors take the non-relativistic limit
in the argument of the energy-conserving $\delta$-function,
$\delta\left( p_0 - M - \varepsilon_D + {\bf p}^2/2M \right)$,
where $\varepsilon_D \equiv M_D - 2 M$ is the deuteron
binding energy.
Furthermore, in order to make a comparison of our full result
with previous calculations \cite{WOL}, we have included
in Eq.(\ref{Dfy}) the ``flux factor'' $(1+p_z/M)$,
which was introduced in Ref.\cite{WOL} by analogy with
unpolarized structure functions.
However, this prescription also differs according to various
authors \cite{FS,CIOFI,KAP}.
On the other hand, the full, relativistic result in
Eq.(\ref{g1Dfull}) is exact (within the impulse approximation)
and contains no such ambiguities.

The three non-factorizable corrections in Eq.(\ref{comps})
are proportional to powers of $|{\bf p}| / M$.
The first two corrections,
\begin{eqnarray}
\delta^{({\cal A})} g_{1D}(x)
&=& {1 \over 8 \pi^2 M_D}
    \int {dy \over y} dp^2
    \left[ {\cal A}^+(y,p^2)
    \right.                                             \nonumber\\
& & \left. \hspace*{3cm}
         - { 1 \over {\cal N} }
           { E_p \over M }
           \left( 1 + { p_z \over M } \right)
           {\cal A}^+_{ON}(y,p^2)\
    \right]
    g_{1N}\left({x \over y}\right),             \label{delA}
\end{eqnarray}
and
\begin{eqnarray}
\delta^{(N)} g_{1D}(x)
&=& {1 \over 8 \pi^2 M_D}
    \int {dy \over y} dp^2\
    {\cal A}^+(y,p^2)\
    g_{1N}^{off}\left({x\over y},p^2\right),    \label{delN}
\end{eqnarray}
arise from the off-shell components associated with the
deuteron--nucleon vertex and the nucleon structure function,
respectively.
In the latter\
$g_{1N}^{off}(x/y,p^2) \equiv \widetilde g_{1N}(x/y,p^2) - g_{1N}(x/y)$.
The $\delta^{(G)}$ correction is given by:
\begin{eqnarray}
\delta^{(G)} g_{1D}(x)
&=& -\ {M_D M \over 2}
       \int dy\ dp^2\
       \left[ \left( M_D - 2 E_p \right)
              { p_z \over M }
              \left( u - { w \over \sqrt{2} } \right)^2
           -\ {3 M_D p_z \over 2 M } v_t^2\
       \right.                                          \nonumber\\
& & \hspace*{0cm} \left.
           +\ \sqrt{6} \left( M_D - 2 E_p \right) \cos\theta
              \left( u - {w \over \sqrt{2}} \right) v_t
       \right]\ p\cdot q\ G_{(p)}(p,q).                       \label{delG}
\end{eqnarray}

To estimate the size of the relativistic corrections
requires a model of the nucleon functions $G_{(i)}$.
While the scaling behavior of $G_{(i)}$
can be derived from the parton model, their complete
evaluation requires model-dependent input
for the non-perturbative parton--nucleon physics.
For this purpose it is useful to analyze the problem
in terms of relativistic quark--nucleon vertex functions,
as described in Refs.\cite{MSTD,MST}.
We use an ansatz in which a suitable set of spin $S=0$ and $S=1$
vertex functions is chosen firstly to parametrize the
unpolarized valence nucleon (viz. $u_V + d_V$ and $d_V / u_V$)
and deuteron ($u_V^D + d_V^D$) data.
This fixes the normalization and momentum dependence of
the vertex functions.
The same set is then used to obtain the $S=0$ and 1
polarized valence distributions
$\Delta q_0$ and $\Delta q_1$.
A simple way to relate these to the polarized distributions
$\Delta u_V$ and $\Delta d_V$ is via the SU(4) spin-flavor
symmetric relations \cite{MM}:\
$\Delta u_V = 3 \Delta q_0/2 - \Delta q_1/6$\ and
$\Delta d_V = - \Delta q_1/3$.
Of course our formal results do not rely on the use of SU(4)
symmetry --- these relations merely provide a convenient way
to parametrize the polarized quark distributions.
Indeed, to reproduce the correct large-$x$ behavior of the
proton and neutron structure functions requires that the
various spin-flavor quark distributions have different
asymptotic $x$-dependence \cite{CT,BBS},
which necessarily breaks SU(4) symmetry.
Within the present approach this is achieved by using
different Dirac structures and momentum dependence
for the $S=0$ and $S=1$ vertices.

We find the polarized and unpolarized data can be fitted
with the structures
$I \Phi^a_{0}(p,k)$ and $\not\!k \Phi^b_{0}(p,k)$
for the scalar vertex,
and $\gamma_5 \gamma_{\alpha} \Phi_1(p,k)$ for the
pseudovector, where $k$ is the quark momentum.
The momentum dependence in the
vertices is parametrized by the simple form:\
$ \Phi(p,k)
= N(p^2)\ \cdot\ k^2 / (k^2 - \Lambda^2)^{n}$,
with $N(p^2)$ determined through baryon number conservation.
A best fit to the experimental nucleon distributions at
$Q^2 = 10$ GeV$^2$ (when evolved from $Q^2_0 \simeq$ (0.32 GeV)$^2$
using leading order evolution\footnote{
While a next-to-leading order analysis is important
for a precise determination of the $Q^2$-dependence of $g_1$
and the Bjorken sum rule \cite{Q2},
the present treatment is perfectly adequate for the purpose
of evaluating the relative sizes of the nuclear corrections.})
is obtained for cut-off parameters
$\Lambda^a_{0} = 1.0$ GeV and
$\Lambda^b_{0} = 1.1$ GeV,
and exponents $n^a_{0} = 2.0$ and $n^b_{0} = 2.8$,
for the two scalar vertices.
These contribute to the total scalar distribution as:
$\Delta q_0(x) = r\ \Delta q^a_{0}(x)
               + (1 - r)\ \Delta q^b_{0}(x)$,
with $r = 0.15$.
The parameters for the pseudovector vertex are
$\Lambda_1 = 1.8$ GeV and
$n_1 = 3.2$.
The mass parameters associated with the intermediate spectator
states are taken to be $m_{0(1)} = (p-k)^2 = 0.9 (1.6)$ GeV.

With these parameters the first moments of the polarized
valence distributions in the proton are
$\int_0^1 dx\ \Delta u_V(x) = 0.99$ and
$\int_0^1 dx\ \Delta d_V(x) = -0.27$,
which saturates the Bjorken sum rule:
$ \int_0^1 dx\ (\Delta u_V(x) - \Delta d_V(x))
= g_A $.
The $x$-dependence of the polarized proton
structure function
$ x g_{1p}(x)
= x (4 \Delta u_V(x) + \Delta d_V(x)) / 18$\
is shown in Fig.1.
In the valence quark dominated region ($x > 0.3$) the
result agrees rather well with the SLAC, EMC and SMC
proton data \cite{SLAC,EMC,SMCP}.
A negatively polarized sea component at $x < 0.3$ would
bring the curve even closer to the data points.

For the deuteron, the structure function calculated from
Eq.(\ref{comps}) is also shown in Fig.1
(scaled by a factor 1/2).
The agreement with the SMC data \cite{SMCD} in the valence
region is also quite good.
Because the structure function is not very sensitive to
explicit $p^2$-dependence in the quark--nucleon vertex functions,
we take $N(p^2)$ to be constant.
The numerical values of these normalization constants
are fixed through valence quark number conservation in
the deuteron to be (1.2, 0.6, 0.8)\% smaller
(for the $\Phi_0^a$, $\Phi_0^b$, $\Phi_1$ vertices respectively)
than in the case of the free nucleon.

The resulting ratio, $g_{1D}/g_{1N}$, is displayed in Fig.2.
For large $x$ it exhibits the same characteristic shape
as for the (unpolarized) nuclear EMC effect, namely a dip
of $\sim$ 7--8\% at $x \approx 0.6$ and a steep rise due to
Fermi motion for $x > 0.6$.
For small $x$ it stays below unity, where it can be reasonably
well approximated by a constant depolarization factor,
$1 - 3/2\ \omega_D$, as is typically done in data analyses
\cite{SMCD}.
Also shown in Fig.2 is the ratio of the convolution ansatz
(Eqs.(\ref{comps}) -- (\ref{DSp})) to the full calculation
(dashed curve).
As one can see, this ansatz works remarkably well for most $x$,
the only sizable deviations occurring for $x > 0.8$.

To obtain a better idea about the precise origin of the
off-shell effects, we plot in Fig.3 the individual off-shell
corrections defined in Eqs.(\ref{delA})--(\ref{delG}),
relative to $g_{1D}$.
For $x < 0.8$ each of the corrections is of order 0.5\% or less.
As $x \rightarrow 1$, however, the convolution-breaking
off-shell effects increase by more than an order of magnitude,
and will need to be accounted for if one is to obtain accurate
information on the $x \rightarrow 1$ behavior of the neutron
structure function extracted from deuterium data.
We have also made the calculation for the case of pseudoscalar $\pi N$
coupling \cite{BG}, where we find the total off-shell correction
to be roughly twice as large \cite{MPT}.
However pseudoscalar coupling is usually considered to be
less realistic than the pseudovector form.
The corrections may also depend on the model for the
off-shell nucleon structure function, although the fact
that the off-shell effects become large at $x \sim 1$
is largely independent of the $x \rightarrow 1$ structure
of the on-shell nucleon input \cite{MPT}.

Finally, as an estimate of the uncertainty, $\Delta g_{1n}$,
introduced into the extracted $g_{1n}$ through neglect of
various nuclear effects, we plot in Fig.4 the quantity
$\Delta g_{1n}(x)
\equiv 2 g_{1D}(x)
       \left( \widetilde R(x) - R(x) \right) /
       \widetilde R(x) R(x)$,
where $R(x) = g_{1D}(x)/g_{1N}(x)$ is obtained in the full
model, while $\widetilde R(x)$ is determined through the
convolution formula (solid) and via the ansatz
$\widetilde R(x) = 1 - 3/2\ \omega_D$ (dashed),
with $\omega_D = 4.7\%$ \cite{BG}.
Neglecting off-shell corrections then leads to an uncertainty
$|\Delta g_{1n}| \sim 0.002$ in the absolute value of $g_{1n}$,
which is still quite small.
For comparison we also illustrate the error associated
with using a different value for $\omega_D$ in
$\widetilde R(x) = 1 - 3/2\ \omega_D$,
namely 5.8\% (dot-dashed) as used in the
SMC analysis \cite{SMCD}.
A larger value for $\omega_D$ would have the effect
of shifting the constant curve in Fig.2 down below
the ``full" result, thereby changing the overall sign
of $\Delta g_{1n}$ in Fig.4.
Thus, depending on the precise value of $\omega_D$ used,
the procedure of applying a constant depolarization factor
may lead to an absolute uncertainty
$|\Delta g_{1n}| \sim 0.01$, or about 10\% of the value
of $g_{1n}$ at $x \approx 0.1-0.2$.

In summary, we have calculated the polarized deuteron structure
function $g_{1D}$ within a covariant framework.
Our analysis includes, for the first time, a detailed investigation
of effects associated with the off-shell structure of bound nucleons
--- in addition to the more familiar Fermi motion and binding effects.
The conventional convolution model can only be recovered from
the full result by taking on-shell limits in the virtual
nucleon structure function and in the nucleon--deuteron
interaction.
In the end, our conclusion is that the major uncertainty
in the extraction of $g_{1n}$ comes from the lack of knowledge
of the deuteron $D$-state wavefunction.
The off-shell corrections for $x < 0.7$ are small, of order
0.5\% out of a total nuclear effect of $\sim$ 7\%.
At larger $x$ the correction increases rapidly to $> 5\%$,
and will be relevant for higher moment analyses
of $g_{1n}$.
Although at present the nuclear effects still lie within
the error bars of available data, in the upcoming,
high-statistics SLAC E154 and HERMES experiments
a careful analysis of all nuclear effects will be
necessary.

\vspace*{1cm}
{\bf Acknowledgements.}

We would like to thank S.Kulagin and H.Meyer
for helpful comments, and W.Weise for a careful reading of the
manuscript.
W.M. would like to thank the theoretical physics group at the
University of Adelaide for
its hospitality during a recent visit.
This work was supported by
the Australian Research Council
and the BMFT grant 06 OR 735.


\newpage

FIGURE CAPTIONS.

1. Valence component of the proton (solid) and deuteron (dashed)
   $g_1$ structure functions at $Q^2=10$ GeV$^2$.
   The data are from Refs.\protect\cite{SMCD,SLAC,EMC,SMCP,E143P}.
   The deuteron data are scaled by a factor 1/2.

2. Ratio of deuteron and nucleon structure functions in the
   full model (solid), and with a constant depolarization factor
   $1 - 3/2\ \omega_D$ (dotted), with $\omega_D = 4.7\%$ \protect\cite{BG}.
   Dashed curve is ratio of $g_{1D}$ calculated via convolution
   and in the full model.

3. Nucleon off-shell corrections to $g_{1D}$:\
   $\delta^{(N)} g_{1D}$ (dotted),
   $\delta^{({\cal A})} g_{1D}$ (dashed),
   $\delta^{(G)} g_{1D}$ (dot-dashed)
   and the sum (solid), as a fraction of the total $g_{1D}$.

4. Estimate of the error, $\Delta g_{1n}$ (scaled by a factor 10),
   introduced into $g_{1n}$ extracted from $g_{1D}$ by neglecting
   off-shell effects (solid), and by using a constant
   depolarization factor $(1-3/2\ \omega_D)$,
   with $\omega_D =$ 4.7\% \protect\cite{BG} (dashed)
   and 5.8\% \protect\cite{SMCD} (dot-dashed).
   To show the significance of the correction,
   $\Delta g_{1n}(\times 10)$ is compared with the
   SLAC E142 $g_{1n}$ data \protect\cite{E142}.

\end{document}